\documentclass[%
 reprint           , 
 superscriptaddress,
 altaffilletter    ,
 nofootinbib,
 aps               ,
]{revtex4-1}

\usepackage{graphicx}
 \graphicspath{{graphics/}}

\usepackage{multirow}
\usepackage{dcolumn} 

\usepackage{amsmath}
\usepackage{amssymb}
\usepackage[version=3]{mhchem}
\usepackage{siunitx}

\usepackage[hidelinks,linktocpage,colorlinks]{hyperref}
\hypersetup{
            colorlinks,
            linkcolor={red!70!black}  ,
            citecolor={green!70!black},
            urlcolor ={blue!70!black}
           }

\usepackage{comment} 
\usepackage[switch]{lineno} 
\usepackage[usenames,dvipsnames]{xcolor}
\usepackage{xspace}

\newcommand{\al}    {\ensuremath{\alpha}\xspace}

\newcommand{\gm}    {\ensuremath{\gamma}\xspace}

\newcommand{\dtest} {\ensuremath{d_\mathrm{test}}\xspace}

\newcommand{\GAGG}  {GAGG\xspace}
\newcommand{\Am}    {\ce{^{241}Am}\xspace}
\newcommand{\Na}    {\ce{^{22}Na}\xspace}

\newcommand{\keVee} {keV$_\mathrm{ee}$\xspace}

\newcommand{\QF}   {QF\xspace}

\newcommand{\uniMIB} {\affiliation{Department of Physics, University of Milano-Bicocca, Milano I-20126, Italy}}
\newcommand{\infnMIB}{\affiliation{INFN -- Sezione di Milano-Bicocca, Milano I-20126, Italy}}
\newcommand{\uniPV}  {\affiliation{University of Pavia, Pavia I-27100, Italy}}
\newcommand{\infnPV} {\affiliation{INFN -- Sezione di Pavia, Pavia I-27100, Italy}}

\begin{document}

 \title{A custom experimental setup for scintillator characterization: \\application to a Ce-doped GAGG crystal}

 \author{L.\ Gironi}   \uniMIB \infnMIB
 \author{S.\ Dell'Oro}\email{stefano.delloro@unimib.it} \uniMIB \infnMIB
 \author{C.\ Gotti}            \infnMIB
 \author{N.\ Manenti}  \uniPV  \infnPV
 \author{E.\ Mazzola}  \uniMIB \infnMIB
 \author{M.\ Nastasi}  \uniMIB \infnMIB
 \author{D.\ Peracchi} \uniMIB \infnMIB

 \date{\today}

 \pacs    {}

\begin{abstract}
 Scintillators are widely used in radiation detection, with continuous advancements enhancing their performance and developing new materials.
 This study presents a custom experimental setup for the characterization of crystal scintillators under different temperature and pressure conditions. The setup is flexible and capable of providing prompt feedback, which is crucial for material development.
 We tested the setup with a \ce{Ce}-doped \ce{Gd_3Al_2Ga_3O_{12}} (\GAGG) scintillator, evaluating its response to different types of radiation, particularly alpha particles.
 These results contribute to a deeper understanding of \GAGG’s scintillation properties, including light output, quenching factors, and pulse-shape discrimination capabilities.
 \\[+9pt]
 Published on: Eur.\ Phys.\ J.\ Plus {\bf 140}, 415 (2025) \hfill DOI: \href{https://doi.org/10.1140/epjp/s13360-025-06350-9}{10.1140/epjp/s13360-025-06350-9}
\end{abstract}

\maketitle

\section{Introduction}

 Solid-state scintillators have long been among the most widely used radiation detectors.
 Thanks to their ease of use and strong performance in spectroscopy and timing measurements, they are commonly employed in a broad range of applications, from fundamental to applied physics.
 Despite this long-standing role in radiation detection, the extremely large interest from the different communities continue to drive extensive research into solid-state scintillators, aimed both at boosting their performance and at enhancing their capabilities.
 Recent progress in material science have already seen the development of innovative scintillating materials, and are now paving the way towards next-generation scintillators.
 Notable examples of innovations include scalable industrial methods for the synthesis of various nanocrystals, which can be embedded in solid (or liquid) host matrices to create novel classes of scintillators, as well as the exploration of hybrid scintillators and other cutting-edge approaches~\cite{Singh:2024acsnano, Dujardin:2018IEEE, CarrDelgado:2024Nano}.

 Conversely, the constant appearance of new materials requires effective approaches for the study of their properties in order to investigate their suitability to be employed as radiation detectors.
 A crucial study in this regard is that of the response of the scintillator to different types of radiation.
 It has been well established that highly ionizing particles produce less light in a scintillating material than electrons of the same energy~\cite{Birks:1951boa}.
 On the one side, this feature can be profitably exploited to perform pulse-shape discrimination analysis in order to clean the signal set from unwanted background events.
 On the other, it implies that when a scintillator is calibrated using \ce{e^-} or \gm sources, ion signals appear at lower energies than their true values. Quantitatively understanding such behavior is crucial for rare-decay and dark-matter searches.
 For the former, this is important to get a correct energy reconstruction, especially of \al decays, e.\, g.\ \ce{^{180}W} in \ce{CdWO_4}~\cite{Arnaboldi:2010tt}, \ce{^{209}Bi} in \ce{BGO}~\cite{Beeman:2012prl} or \ce{^{204}Pb} in \ce{PbWO_4}~\cite{Beeman:2012wz}.
 For the latter, it is crucial to identify and delimit the region where to look for the WIMP-induced signals~\cite{Strauss:2014zia,Joo:2018hom}.

 In this article, we introduce a custom-designed setup for the characterization of crystal scintillators.
 The setup is flexible and well-suited to providing prompt feedback during the development of new materials.
 It can be easily adapted for various studies of the crystal properties, allowing characterization under different temperature and pressure conditions and accommodating studies with different types of radiation sources.
 We also present the results of a first case study on a GAGG sample. These results confirm findings already reported in the literature, while extending the understanding of this scintillator, particularly regarding its response to \al particles at low energies, which we studied by varying the air pressure in the setup while keeping the crystal temperature and the source distance constant.
 
\section{Experimental setup}

\begin{figure}[tb]
  \includegraphics[width=.7\columnwidth]{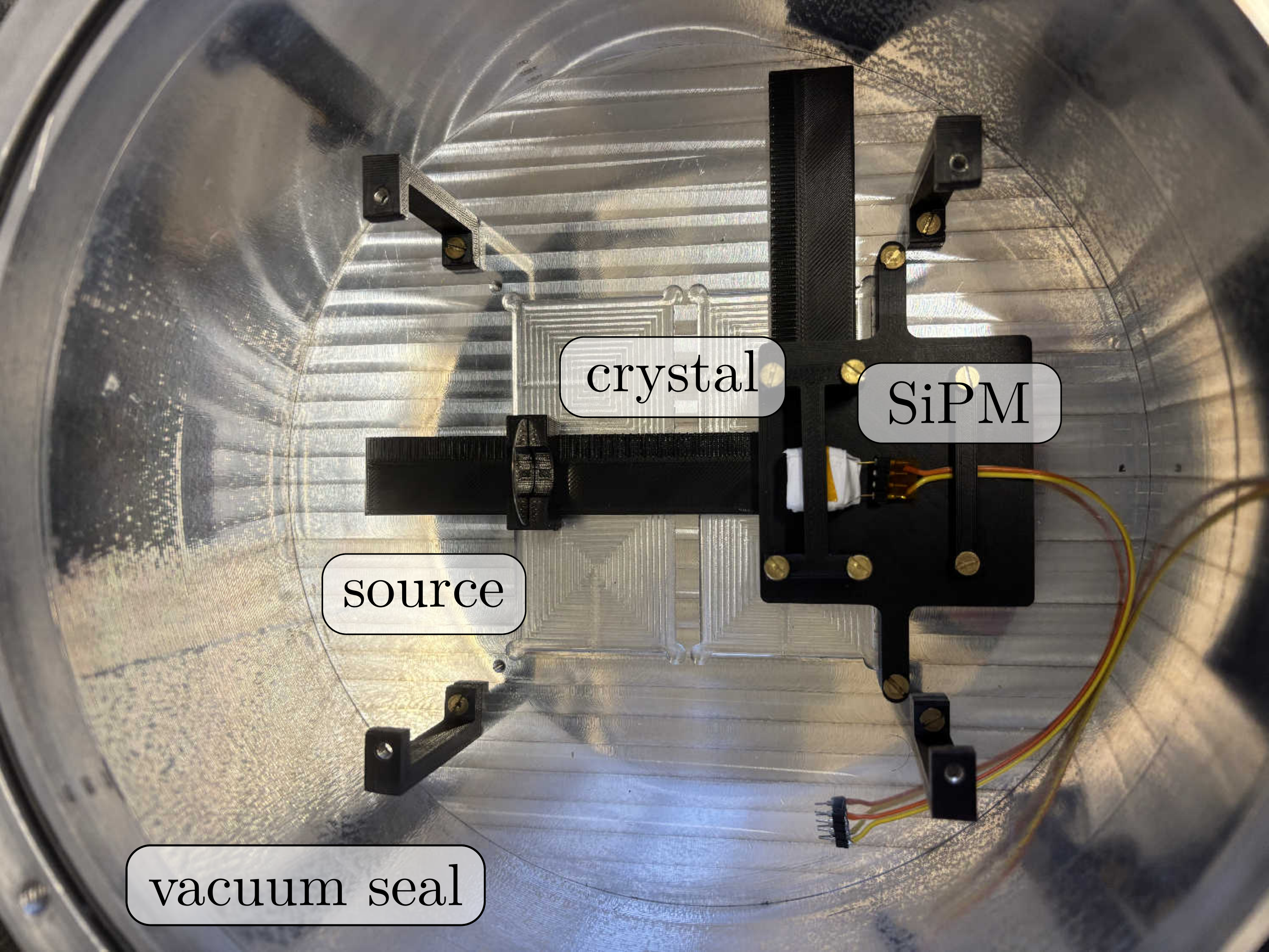}
  \caption{Top-view of the experimental setup.
           A crystal coupled to a SiPM is exposed to an \al source.}
  \label{fig:setup}
 \end{figure}

 The crystal scintillators are coupled to Silicon Photomultipliers (SiPM) to readout the light.
 These detectors are operated inside an aluminum vacuum-tight cylindrical chamber ($25$-cm diameter, $10$-cm height) that is able to stand pressure in the range $10^{-2}$ mbar -- $1.2$ bar.
 The chamber is instrumented with $4$ Peltier cells that allow us to operate a detector at a temperature comprised between $-10~^\circ$C and $30~^\circ$C ($\pm 2~^\circ$C). The excess heat generated is removed by a water-cooling circuit running inside the support plate of the chamber, to which the Peltier cells are thermalized (Fig.~\ref{fig:setup}).

 The SiPMs are S13360-series Multi-Pixel Photon Counters by Hamamatsu Photonics, with a sensitive area of $6\times 6$ mm$^2$ ($14400$ single cells with $50$-$\mu$m pixel pitch).
 The SiPM cathode is biased with a positive voltage through a resistor from a bench top power supply; the anode is connected to ground.
 The current signals from the SiPM cathode are AC-coupled to a transimpedance amplifier (AD811).
 The amplifier was chosen for its wide bandwidth and low voltage noise, and is able to drive a $50$-$\Omega$ terminated line with a good dynamic range before saturating. The board is designed with four independent channels, but only one channel has been used in this work.
 The data acquisition and digitalization is performed  by a Rohde\&Schwarz MXO4 oscilloscope, which directly records and stores the signal waveforms; the acquisition parameters can be set remotely.
 Finally, the waveform are processed offline by a python-based software that computes the pulse parameters (amplitude, rise and decay time, \dots), performs selection cuts, and generate and calibrate the energy spectra.

\section{Case study: \GAGG crystal}

 The first application (and validation) of our setup has been the characterization of a \ce{Ce}-doped \ce{Gd_3Al_2Ga3O_{12}} (\GAGG, \cite{Kamada:2011_GAGG}) scintillator.
 \GAGG is a fairly novel material suitable for particle counting and spectroscopy given its fast timing and high light output.
 The response of \GAGG to different types or radiation has already been investigated~\cite{Iwanowska:2013_GAGG,Taggart:2019_GAGG_n,Furuno:2021cns}, though not fully explored.

 We coupled a $10\times10\times10$ mm$^3$ \GAGG-T sample from Epic Crystal to a SiPM.
 In order to maximize the light collection, we coated with PTFE tape the lateral faces of the crystal, while the face opposite to the SiPM was left bare so that it could be exposed to \al particles.
 We anchored the crystal to a support structure inside the chamber; for this work we did not use the Peltier cells.

 As a preliminary study, we verified the linearity of the detector response by using a set of commercial sources that provide standard-candle full-energy peaks, namely \ce{^{241}Am} ($59.5$ keV), \ce{^{22}Na} ($511.0$ keV), \ce{^{137}Cs} ($661.7$ keV).
 This series of energies fully covers the range of interest for \gm-ray events given the dimensions of the crystal. Likewise, \al events depositing a few MeV in the crystal lie in the same portion of the spectrum, due to the quenching for \al events in scintillators (Sec.~\ref{sec:quenching}).
 The \ce{^{241}Am} source also provides a monochromatic \al peak at $5.49$~MeV, whose energy deposited in the crystal depends on the distance and on the pressure in the experimental chamber. 
 
\subsection{Light Output}
\label{sec:light_output}

 \begin{table}[tb]
  \centering
  \caption{Parameters, decay constants and fraction of the short time component, extracted from the reconstruction of particle-physics pulses with the \GAGG scintillator.}
  \vspace{5pt}
  \setlength{\tabcolsep}{8pt}
  \begin{tabular}{lrr}
   \toprule \\[-8pt]
   Parameter          &Value &Error       \\[2pt]  
   \hline   \\[-7pt]
   $\tau_\mathrm{short}$ [ns]           &$95$       &$5$    \\[5pt]
   $\tau_\mathrm{long}$ [ns]            &$345$      &$10$   \\[5pt]
   $F_\mathrm{short}$ (\al)             &$0.4$     &$0.1$ \\[5pt]
   $F_\mathrm{short}$ (\gm)             &$0.7$     &$0.1$ \\[5pt]
   \toprule
  \label{tab:pulse_reconstruction}
  \end{tabular}
 \end{table}
 
\begin{figure}
  \includegraphics[width=1.\columnwidth]{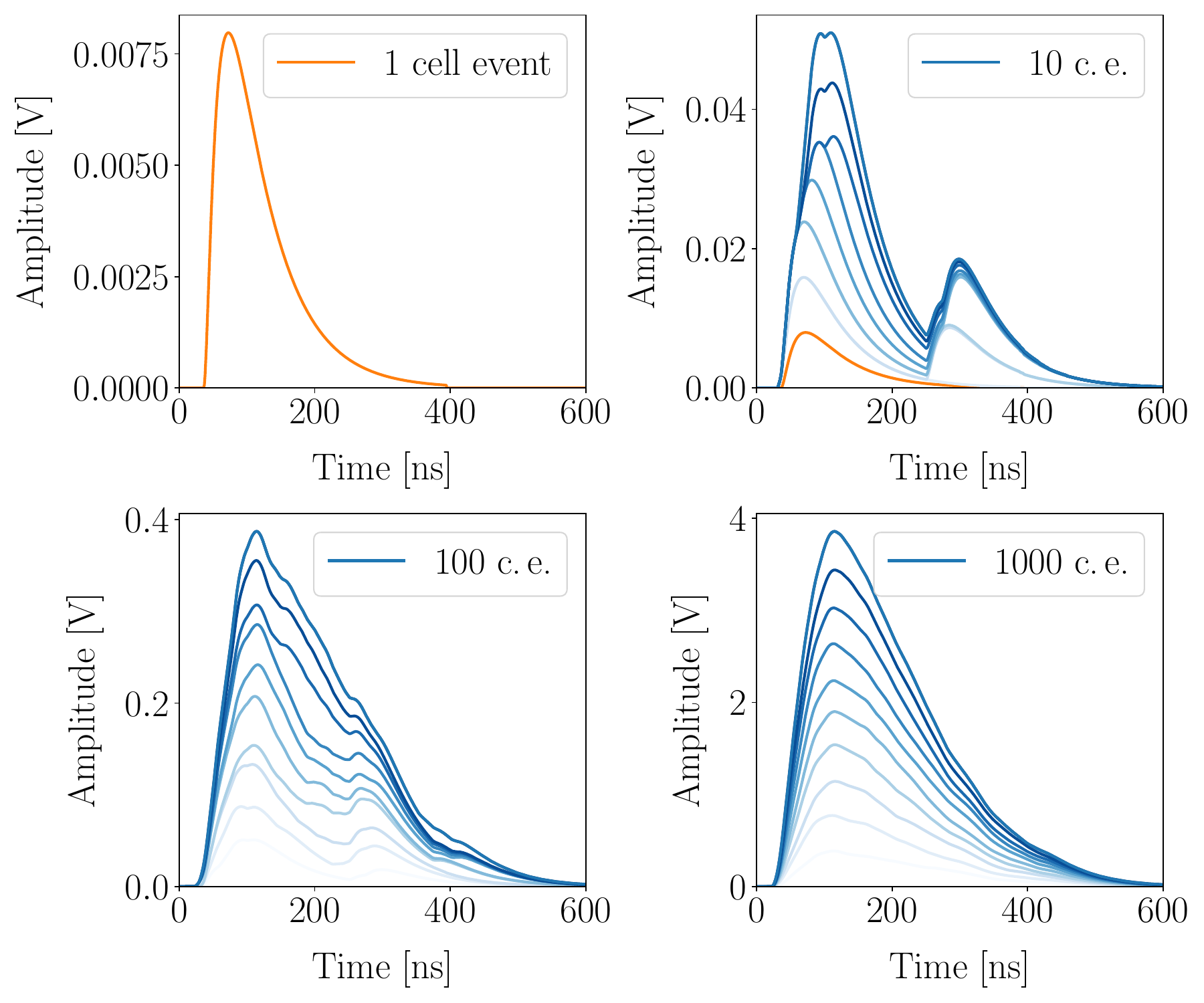}
  \caption{Construction of a particle-physics-like pulse starting from a single-cell event from the SiPM (orange pulse).
           The blue pulses are constructed by summing an increasing number of single-cell events, with the gradient representing intermediate stages of the buildup.
           Unlike the description in the text, only one time constant is used here for illustration purposes, allowing a clearer comparison between the single-cell event and the particle-physics-like pulse by keeping the same time scale.}
  \label{fig:single_cell}
 \end{figure}

 We constructed a template pulse from individual SiPM single-cell signals, which we could easily obtain from the dark counts, and used it as a building block to model particle-induced pulses from both \gm and \al events (Fig.~\ref{fig:single_cell}). The pulse shape was reproduced by summing a sufficient number of these single-cell templates, randomly generated according to a two-exponential distribution, with a short and a longer decay-time constant, which can model the time distribution of photons from most scintillators.
 For \gm events, the reference pulse was obtained by averaging over a population of $511$-keV interactions from the \ce{^{22}Na} source. The reference pulse for \al events was derived from $5.49$-MeV interactions from the \ce{^{241}Am} source, with the chamber pressure adjusted to shift the peak close to $511$ keV electron-equivalent (\keVee).

 To extract the time constants, the relative weights of the two components, and the number of single-cell events, we used a minimization algorithm to compare the constructed particle-physics-like pulses with the reference pulses.
 In principle, the decay constants and relative weights could have been obtained by directly fitting a \gm (\al) pulse. However, our approach has the advantage of naturally incorporating the system response, which would otherwise need to be included in the fit function by convolving it with the scintillation process.
 The values we obtained for the \GAGG parameters are reported in Table~\ref{tab:pulse_reconstruction}.
 In particular, the values we extracted for the time constants are compatible with those found in Ref.~\cite{Furuno:2021cns} and those provided by the crystal producer, and the same holds for the relative contribution in the case of \gm events.
 
 The number of single-cell events equivalent to a certain amount of deposited energy depends not only on the intrinsic properties of the scintillator, but also on the experimental configuration.
 Starting from the amount of firing cells, the SiPM producers provide an empirical formula to retrieve the number of photons that reached the sensor~\cite{HPK:web_formula}.
 The latter is obtained based on the hypothesis that, given the specifications of the SiPM, each cell can either fire once or more depending on the pulse width.
 In turn, from the number of photons reaching the sensor it is possible to get an estimate of the \GAGG light yield, although approximated.
 
 We found that with our setup around $1150$ single-cell pulses correspond to an \ce{e^+e^-} annihilation event of $511$~keV, i.\,e.\ about $2250$ per MeV deposited in the crystal.
 By using different sources, we could actually verify the linearity between the amplitude of the \gm events and the number of single-cell pulses.
 These $2250$ single-cell pulses per MeV, according to Ref.~\cite{HPK:web_formula}, correspond to $\sim 6700$ photons per MeV. Considering that this refers only to the active area of the SiPM, they become $\sim 18600$ over the whole crystal face to which the sensor is coupled. In addition, since the crystal face opposite to the SiPM was bare, we can conservatively add another factor $2$ to the computation.
 We thus obtain an estimate for the \GAGG light yield of $\sim 37200$ photons per MeV, fairly compatible with the nominal value of $\sim 40000$ indicated by the crystal producer, especially considering that we did not include minor effects such as some light dispersion on the coated side faces of the crystal, which still remains unavoidable.
 
\subsubsection{Pulse-shape discrimination}
 
 \begin{figure}
  \includegraphics[width=1.\columnwidth]{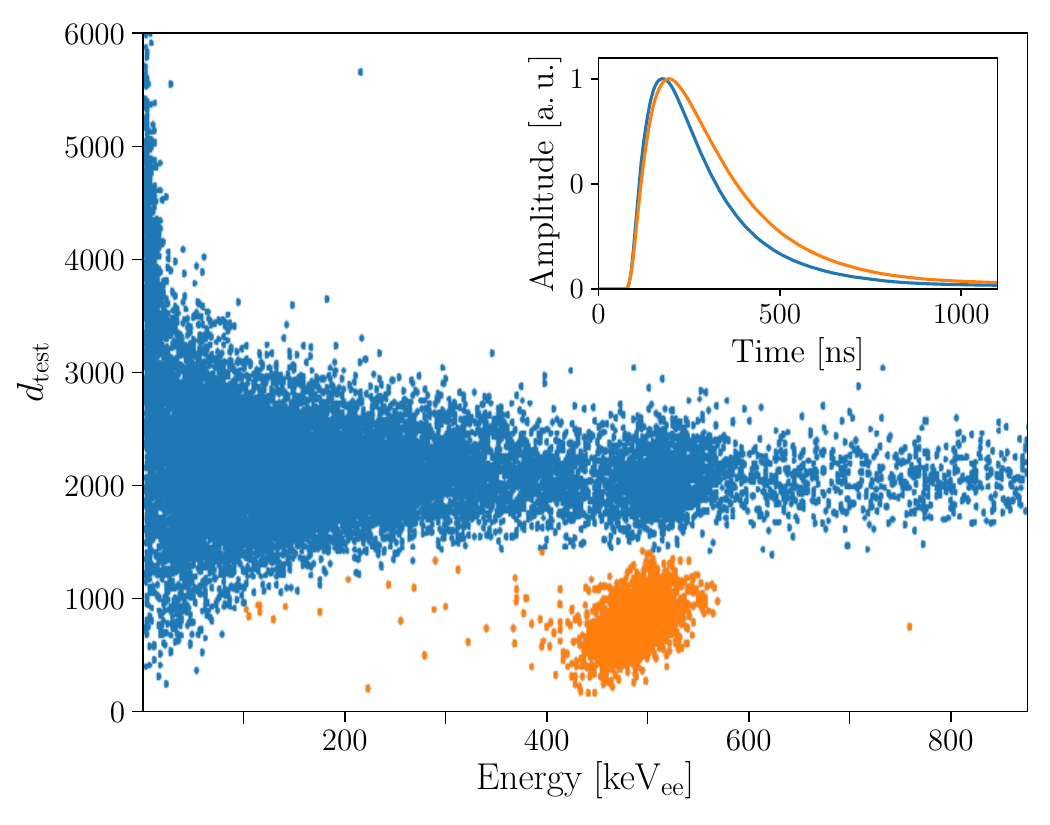}
  \caption{Scatter plot of \dtest as a function of the pulse energy.
           The population of \al events from the $5.49$-MeV peak from \Am (orange) can be easily clustered against the \gm continuum and the $511$-keV peak from \Na (blue).
           On the top right corner, the normalized reference pulses for the two populations are reported.}
  \label{fig:scatter}
 \end{figure}

 \GAGG exhibits particle discrimination capabilities due to differences in light emission timing between \al and \gm interactions.
 As shown in Table~\ref{tab:pulse_reconstruction}, \al-induced scintillation occurs on a different timescale than \gm-induced scintillation, resulting in distinct waveform shapes that correlate with the type of particle interaction.
 By leveraging this property, pulse-shape discrimination can be applied to classify events, isolate specific interaction families, or suppress the \gm background in \al spectrum analysis.
 To quantify this difference, we define a test parameter:
 \begin{equation}
  \dtest \equiv \sum_{k=1}^N {\frac{(k - k_\mathrm{ref})^2}{\sigma^2}}.
  \label{eq:scatter}
 \end{equation}
 
 The distance \dtest compares each point $k$ of a waveform to the corresponding one of a reference \al pulse, which we built by averaging events from the \Am $5.49$-MeV peak, with weighting based on the baseline error $\sigma$.
 Using \dtest, we apply a clustering approach to determine which reference pulse best matches each waveform.

 We exploited this technique during the measurements with \Am sources described in Sec.~\ref{sec:quenching}.
 As it can be seen from Fig.~\ref{fig:scatter}, the \al events from the $5.49$-MeV are well separated from the \gm populations.
 We verified that by changing the pressure in the chamber, the two populations keep consistent values of \dtest and the only relevant difference in the various measurements is the shift of the \al-event cluster in energy.
 
 \begin{figure*}[tb]
  \centering
  \includegraphics[width=1.\textwidth]{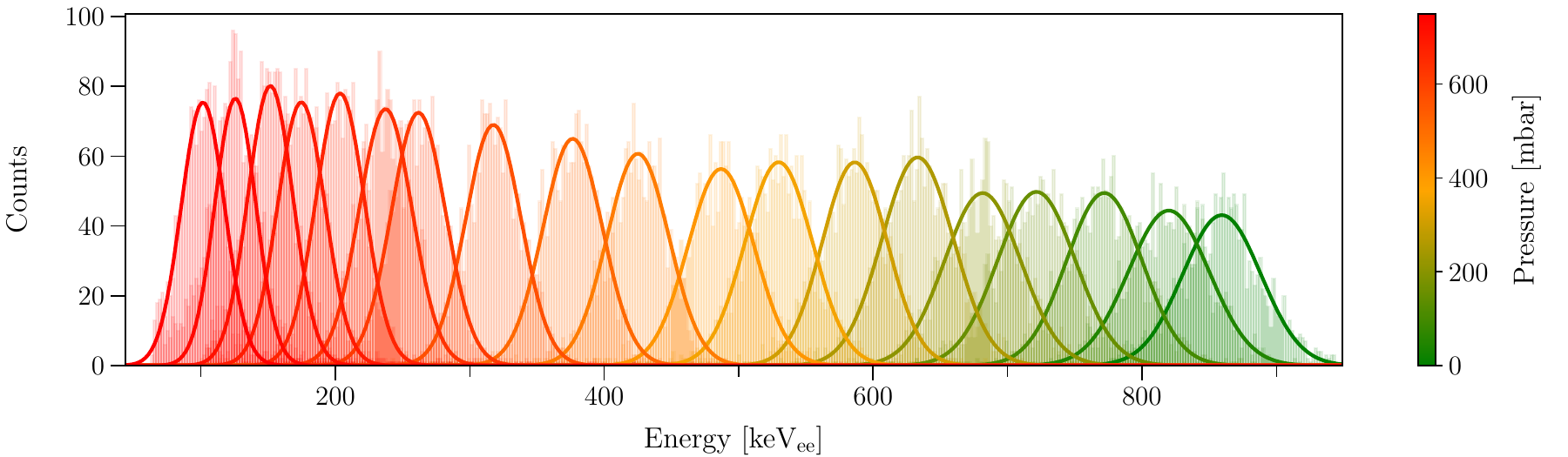}
  \caption{Series of $5.49$-MeV \al peaks from the \Am source taken at different pressure inside the chamber.
           The peak progressively shifts towards lower amplitudes and shrinks, as explained in App.~\ref{app:resolution}.}
  \label{fig:spectra}
 \end{figure*}

\subsection{Quenching factor for \al particles}
\label{sec:quenching}
 
 The light produced in a scintillator by \al particles, and by highly ionizing particles more in general, is lower than that produced by electrons of the same energy.
 This behavior is due to the higher density of charge carriers along the track, which saturate the response of the luminous centers~\cite{Lecoq:2020itu}, and it is commonly parametrized by a \emph{quenching factor} (\QF) indeed defined as the ratio of the light yield of \al/ions to that of electrons of the same energy.
 The \QF depends on the particle energy and has its minimal when the stopping power is maximal; its impact for \al's is thus larger around $1$ MeV in many crystal scintillators~\cite{Tretyak:2009sr}.

 In order to characterize the \QF of \GAGG, we took a series of measurements at increasing pressure, from $\sim2$~mbar to $750$ mbar, while exposing the crystal to the \Na and \Am sources.
 We fixed the distance between the \Am source and the crystal to $51$ mm, this value being the result of a compromise between a sufficiently-high rate from the source and an appreciable effect of the air at low pressures.
 From the different spectra, we could observe that while the \ce{e^+e^-} annihilation peak remains basically unchanged and it is always reconstructed at the same position, the $5.49$-MeV peak from \Am shifts towards lower amplitudes and shrinks as the pressure increases (Fig.~\ref{fig:spectra}, see App.~\ref{app:resolution} for details).
 It is interesting to notice that at low energies, we observed both the $59.5$-keV \gm peak from de-excitation of \ce{^{237}Np} (onto which \Am decays) and, below about $200$ mbar, the continuum due to the electrons from internal conversion.
 
\begin{figure}[tb]
  \centering
  \includegraphics[width=1.\columnwidth]{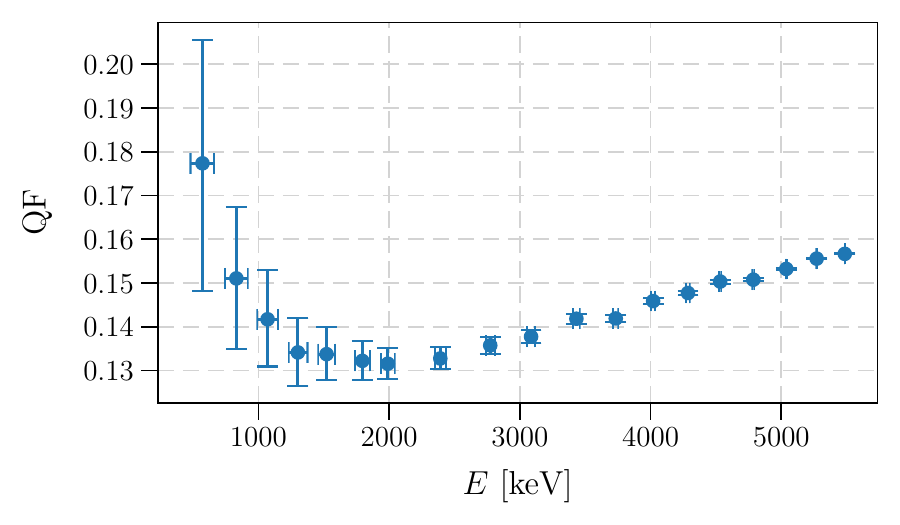}
  \caption{\QF for \al particles in \GAGG rendered as ratio between light yield and energy as a function of the \al energy.
           Values comprised between $\sim 500$~keV and $5.49$ MeV are obtained by taking the mean peak from \Am and varying the pressure in the chamber.
           The errors associated to the measurements are propagated from the uncertainty on the distance between the source and the crystal.
          }
  \label{fig:L_E}
 \end{figure}

 We analyzed the ratio between light yield $L$ and \al energy $E$, as a function of $E$.
 $L$ was estimated from the position of the $5.49$-MeV peak during each measurement, while the $E$ was obtained from Monte-Carlo simulations.
 We used the Geant4~\cite{GEANT4:2002zbu} toolkit to build a replica of our setup at the different pressures.%
 \footnote{The simulations also confirmed the contributions at low energies as ascribable to \Am.}
 From these, we could retrieve the actual energy of the \al particle taking into account the loss due to the interaction with air; to this value we associated a systematic error by slightly varying the source distance in the simulated setup.
 The result is shown in Fig.~\ref{fig:L_E}, where the \QF, or relative light output, is evaluated for energies of the \al particles between $\sim 500$~keV and $5.49$~MeV.
 The values for the \QF range from $\sim 0.18$ to down to $\sim 0.13$, the maximum value being characterized by a quite large error associated to the precision on the distance, despite this was below $2\%$.
 The observed trend follows the description in Ref.~\cite{Tretyak:2009sr}.

\subsection{Summary \& outlook}

 In this work, we present a custom-designed experimental setup for characterizing crystal scintillators and demonstrate its effectiveness through a detailed analysis of a \ce{Ce}-doped \GAGG crystal.
 Our setup, which integrates a SiPM readout system and a vacuum-tight chamber, provides an optimal platform for investigating scintillator properties under various environmental conditions and radiation sources.
 The results from the \GAGG case study confirm a light yield consistent with previous predictions and measurements. Moreover, to the best of our knowledge, they enable the first measurement of the energy-dependent quenching factor for this crystal. Additionally, we demonstrate effective discrimination between \al particles and \gm radiation down to a few hundred \keVee, leveraging pulse shape differences.
 The methodology developed in this work can be readily extended to other scintillators, offering a robust framework for characterizing new scintillating materials in both fundamental and applied physics research.
 
\begin{acknowledgments}
 We thank the mechanical workshop team of the INFN of Milano-Bicocca for the constant and constructive support in the design and construction of the setup.
 This work is performed in the framework of the European Pathfinder Open project UNICORN (GA 101098649) and has been partially supported by the Italian Ministry of University and Research (MUR) through the grant Progetti di ricerca di Rilevante Interesse Nazionale (PRIN grant no.\ 2020H5L338). \\[5pt]
\end{acknowledgments}

  \begin{figure}[t]
  \centering
  \includegraphics[width=1.\columnwidth]{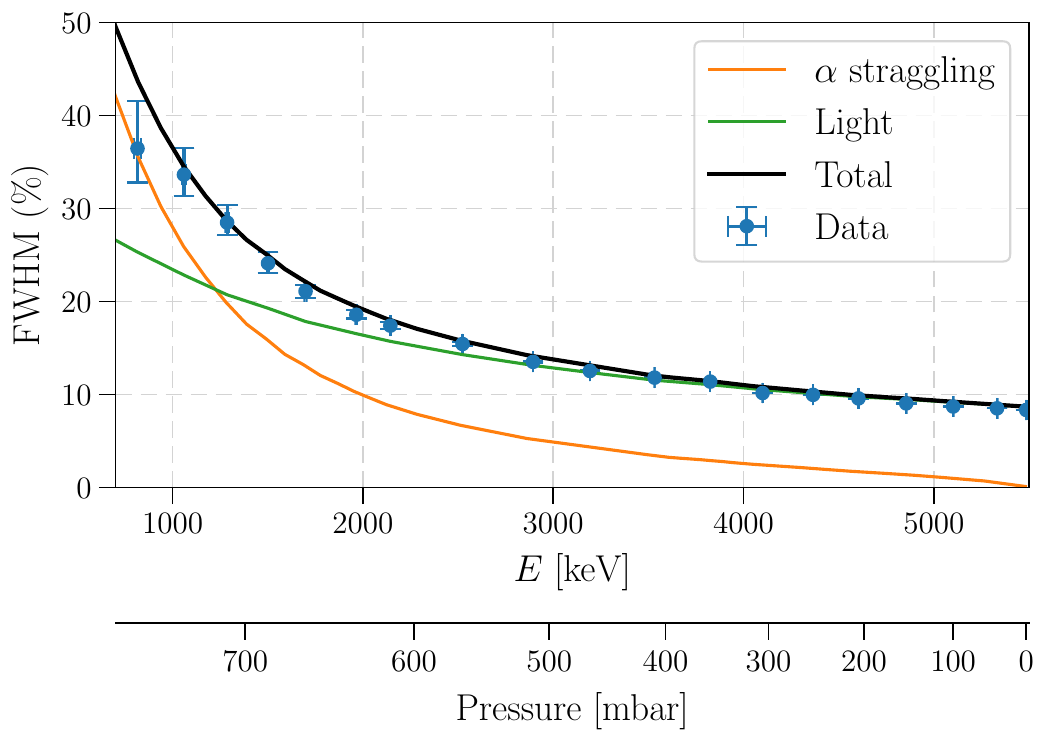}
  \caption{Contributions to the FWHM of the $5.49$ MeV \al peak from \Am as a function of the \al energy and the chamber pressure.
           The source is placed at $51$ mm from the detector.}
  \label{fig:energy_res}
 \end{figure}
  
\appendix

\section{\al-peak energy resolution}
\label{app:resolution}

 As shown in Fig.~\ref{fig:spectra}, both position and width of the $5.49$-MeV peak from \Am vary at different pressures. 
 The interaction of \al particles with air results in an energy loss which affects both features of the \al peak.
 On one hand, the mean energy of the \al particles reaching the crystal decreases as the pressure increases.
 On the other, the reduced energy deposition in the scintillator leads to less light production, thereby worsening the resolution (FWHM) due to the carrier statistics.
 While the former behavior is immediate to appreciate, the latter is a little more subtle to perceive, since it is hidden by the fact that the absolute resolution is actually better at higher than lower pressures.

 Figure~\ref{fig:energy_res}  provides a breakdown of the main factors determining the resolution across different pressures.
 At low pressures, i.\,e.\ at higher deposited energy, the FWHM is primarily affected by the fluctuations in the number of fired cells in the SiPM.
 At high pressures, the energy loss of the \al particles with air becomes more important and the statistic fluctuations of light carriers reaching the crystal dominates the FWHM.
 In particular, the fluctuation in the number of cells fired in the SiPM was estimated by comparing the height of individual cells with the amplitude of the scintillation pulses, while the fluctuation in the \al energy loss was extracted from the Geant4 simulation of the system.

\bibliography{ref} 

\end{document}